# Preparing Millennials as Digital Citizens and Socially and Environmentally Responsible Business Professionals in a Socially Irresponsible Climate

Barbara Burgess-Wilkerson, Clovia Hamilton, Chlotia Garrison, Keith Robbins
Winthrop University

**Abstract**

As of 2015, millennials (born in the 1990's) became the largest population in the workplace – and are still growing. Studies indicate millennials are tech savvy but are lagging in the exercise of digital responsibility. In addition, they are passive towards environmental sustainability and fail to grasp the importance of social responsibility. This paper provides a review of such findings relating to business communications educators in their classrooms. The literature should enable the development of millennials as excellent global citizens through business communications curricula that emphasize: digital citizenship, environmental sustainability and social responsibility. The impetus for this work is to provide guidance in the development of courses and teaching strategies customized to the development of millennials as digital, environmental and socially responsible global citizens.

**Introduction**

Since 2015, millennials are the most represented generational population in the workplace. By 2030, 77 million millennials will make up 75% of the workforce (Fry, 2018). It is likely that the next two decades will bear witness to this new generation confronting issues of global urgency amid demands for public action. One of the greatest challenges facing today's educators is properly equipping the next generation of business leaders with a blueprint for these future challenges. The good news is that millennials possess many admirable attributes. Inarguably millennials are technologically more sophisticated than any previous generation. Millennials are viewed as confident, team-oriented and high achievers (Klass & Lindenberger, 2017). Despite these qualities, a recent national survey determined that barely one adult in three believe millennials as adults will make the world a better place (Strauss, 2003).

Although millennials are the most tech savvy generation of all time, many lack competence in digital responsibility; have a passive attitude towards environmental sustainability and fail to grasp the importance of social responsibility (Davis, 2016; Ribble, 2011). Digital citizenship, environmental sustainability, and social responsibility are key components of" socially responsible ethics." Their definitions follow.

Mike Ribble, author of "Raising a Digital Child" is considered the godfather of digital citizenship and is at the forefront of conceptualizing digital citizenship in schooling. He argues that digital citizenship should be viewed as a way of thinking and whenever possible educators should find ways to incorporate digital citizenship lessons and examples in every classroom. A "responsible digital citizen" is a person skilled in using the internet to buy and sell products/services safely; engages in making practical, safe, responsible, ethical, and legal use of technology; understands the rights and responsibilities that come with being online and is someone who uses technology in a positive manner (Cambridge English Dictionary, 2018; Davis, 2016; Ribble, 2011).

Digital citizenship and environmental responsibility both fall into the over-arching category of social responsibility. Social responsibility or "social responsibility ethics" is an ethical framework that includes individual and organizational obligations to act for the benefit of society and is the duty of every person. Social responsibility refers to a balance between economic development, the welfare of society and the environment. Social responsibility must be embraced by each generation as the actions of one generation will impact the next. Some argue that the ethics of the past have been corrupted and are no longer adequate for the self-governance of businesses and individuals because of changes in the environment, as well as cultural and political norms (Ferrell, 2015).

"Environmental sustainability" is typically defined as the ability to operate today in a way that does not threaten the environment in the future (Foster, 2016, p. 42). Herman Daly, a pioneer in ecological sustainability looked at the problem of environmental maintenance. In 1990 he proposed that an environmentally responsible citizen adheres to ecological sustainability that includes: a) concern for renewable resources - the rate of harvest should not exceed the rate of regeneration (sustainable yield); b) concern for pollution – the rate of waste regeneration should not exceed the assimilative capacity of the environment (sustainable waste disposal); c) concern for renewable resources – the depletion of the nonrenewable resources should require comparable development of renewable substitutes for those resources (Herman E. Daly, 1990; Herman E. Daly, 1990).

Besides environmental sustainability, there are social sustainability and economic sustainability practices. Social sustainability includes product safety, workforce health and safety, ethics and governance, and the improvement of the quality of life in communities. Economic sustainability includes: building high performance organizations; building businesses with sound financial plans; acquiring and managing resources in an effective and efficient manner; and emergency preparedness (Collier, 2007).

**Relevance of Business Communications**

Exploring these topics within the context of business communication is nothing new. Some business communications curricula go beyond the fundamentals of writing and speaking in business to encouragement of critical discourse of events, trends and perspectives impacting how communication takes place in a business setting. Critical thinking and problem-solving are promoted in some business communications classrooms. Barton (1990) argues for the enhancement of business communications courses to extend the course beyond punctuation and style into more of a problem-solving realm. Critical thinking is considered an essential component of managerial literacy; Bloch and Spataro (2014), argue for a two-pronged approach in a business curriculum: a) clearly defining critical thinking and selecting an accessible model for applying it and ; b) integrating critical thinking consistently throughout the business curriculum. Pennell and Miles (2009) argue for problem-based learning in the business communications classroom.

Infusing discourse surrounding digital responsibility, socially responsible ethics and environmental sustainability as part of a critical thinking and problem-solving strategy are part of some business communications curriculum as well. For example, Chaudhri (2014), highlights the importance of corporate social responsibility (CSR) and argues that communication is central to the enactment of socially responsible behavior. He uses CSR scholarship as a basis for understanding how managers construct or articulate the case for communication in CSR, suggesting communication plays an important and multidimensional role. He also highlights the role of the media as an (dis)enabler for "getting the word out." Likewise, Vernuccio (2014) studied the extent to which companies embracing a corporate social responsibility agenda adapt and align their value systems to reflect such commitment by juxtaposing corporate values, corporate social responsibility values and how the configurations of the two impact the corporate communications.

Kim and Lee (2018) highlight the impact of cybersecurity breaches that are creating high-impact crisis for many corporations and how corporations protect their reputations through effective crisis communication. They examined 108 official statements issued in both the U.S. and South Korea when cybersecurity breaches threatened their corporate reputations. The findings indicated the statements varied based upon cultural dimensions that included the communication style (low-context vs. high-context).

Lewis and Speck (1973) conducted a literature review of ethical opinions to help frame the discussion of business ethics to address: 1) why be ethical and, 2). how to be ethical. Barton (1990) argues for the use of case studies for consideration to integrate crisis management into the business communication curriculum. Cases studies also provide opportunities for discourse from several perspectives such as crisis communication, social responsibility, ethics and human rights. "Say It Isn't So, Lady O: A Sex Scandal at the Oprah Leadership Academy for Girls," highlights the allegations of sexual molestation in the Oprah Leadership Academy in Myerton, South Africa providing an opportunity to critically reflect on how leaders communicate and address tough issues in a manner consist with corporate and individual values (Burgess-Wilkerson, Fuller and Frederick 2015).

The purpose of this paper is to provide recommendations based on an analysis of the literature regarding pedagogies that will empower millennials to become excellent global citizens. This research describes how current events are to be used to prepare millennials for professional business communication in the desired capacity as excellent global business citizens. The impetus for this work is to provide guidance in the development of courses and teaching strategies customized to the development of millennials as responsible digital global citizens from an individual, employee and societal perspective.

## Background

Since millennials make up half of the public and private workforce (Raytheon, 2014, 2016, 2017), this study focuses on the potential adverse impact of the present socially irresponsible climate on millennials. Several recent high-profile incidents over the past five (5) years motivated this study. These incidents are listed in Table 1.

Table 1

High Profile Incidents that Motivated the Study

| Year | Industry | Issue | Current Events |
|---|---|---|---|
| 2017 | Retail department store sales | Cybersecurity | The Target store had to pay $18.5 million to 47 states in the United States of America as a settlement for a 2013 security breach. In 2013, hackers obtained credit card numbers, names and other personal information about 40 |

| Year | Industry | Issue | Current Events |
|------|----------|-------|----------------|
|      |          |       | million Target customers. The hackers stole credentials from a third-party vendor that the hackers used to access the customer database. The hackers then installed malware to capture the data (Abrams, 2017). In 2014, Target's CEO Gregg Steinhafel resigned following this massive data breach because the company's board of directors decided it was time for new leadership (Team, 2014). |
| 2016 | US Federal government | Cybersecurity | Further, with regard to computer hacking, it has been declared undeniable that Russia affected the 2016 election (McKew, 2018). An indictment was issued by the United States' head of the Special Counsel investigation of Russian interference in the 2016 United States elections and related matters. Robert Mueller against Russia's Internet Research Agency for information warfare that targeted the American public during the 2016 elections related to issues of use of propaganda and disinformation; and alleged hacking of the election systems and hacking to obtain information illegally. The propaganda is the result of an expenditure of millions of dollars over several years to build a system to influence American opinion with the use of individuals and groups with hidden identities. Their identities were hidden with servers and VPNS used to mask their location; and to launder payments. Paid and promoted ads were not used. Instead data driven targeted social media marketing videos, photos, memes, and text messages were crafted, refined and posted based on data analytics (McKew, 2018). |
| 2018 | Entertainment | Sexual Harassment | The movie producer Harvey Weinstein has been accused by several women of using his power to lure them into hotel rooms where he allegedly committed acts of sexual assault (McKinley Jr, 2018). Actor and comedian Bill Cosby has been convicted of drugging and sexually assaulting women (Bowley, 2018; Thornton, 2018).The hip hop mogul Russell Simmons is facing several accusations of sexual misconduct including harassment, assault and rape.  In November 2017, Simmons stepped down from all of his businesses (Snapes, 2018). In addition, in January 2018, allegations of Las Vegas hotel mogul Steve Wynn were published. There were reports that Wynn sexually harassed, coerced and assaulted Wynn resorts' employees over decades (Richard, 2018). |
| 2015 | Higher Education | Sexual Harassment | A case in point began with the September 2015 investigation of Michigan State University (MSU) by the U.S. Department of Education's Office for Civil Rights. The investigative report stated that MSU failed to address |

| Year | Industry | Issue | Current Events |
|---|---|---|---|
| | | | complaints in a prompt and equitable manner. In August 2016, the Indianapolis Star published an investigative report into the USA Gymnastics and how it handled complaints of sexual abuse. Thereafter, although MSU asked their medical doctor Larry Nassar to step down from his clinical and patient duties, he continued to see patients for 16 months. In December 2017 Nassar was finally convicted and sentenced to 60 years in federal prison for child pornography charges he admitted to (Lacy, 2018). |
| 2017 | Emergency preparedness and relief | Environmental justice; Human Rights | Hurricane Maria and Hurricane Harvey devastated Puerto Rico and the State of Texas in 2017 with historic flooding and wind damage. As of May 2018, half of the Puerto Rico population still does not have power. They are suffering from a dilapidated, corroded and poorly maintained power grid. Their power supplier PREPA and Puerto Rico are both bankrupt (Glanz, 2018). The State of Texas fared better. For example, Toyota partnered with emergency relief nonprofits to plan, assess, train, guide and assist with emergency relief (Evans, 2018). |
| 2018 | US federal government | Human Rights | Facebook, Google, Apple, and the Business Roundtable representing Walmart, General Motors, Boeing, JPMorgan Chase, Mastercard and others spoke out against the US child separations at the Mexican border (Durbin, 2018) These companies wield a great deal of influence and used their influence to speak out against this American policy calling it inhumane. It is important to note that several high technology companies such a Hewlett Packard, Thomson Reuters, Microsoft and Motorola have contracts to develop and implement advanced data analysis and tracking for the U.S. Immigration Customs and Enforcement (ICE) agency. After threatened sanctions by the United Nations, in June 2018, a Presidential executive order was signed to end the inhumane family separations (Hohmann, 2018). |
| 2017 | Information Technology and Automotive | Workplace bullying | Notable CEO bullies who have displayed Machiavellian type cut-throat behavior include, but are not limited to, Larry Ellison at Oracle, Mark Pincus of the gaming company Zynga, and Martin Winterkorn of Volkswagen (Hamilton, 2017). There is a *lack of diplomacy* in communications between employer and employees. |

Thus, it is imperative that business students be educated to be aware of potential c*ybersecurity risks* and the relevance from a business communications perspectiv*e*. They can be taught to critically examine strategies to be socially responsible when delivering negative news and how during a crisis, corporate messaging can simultaneously protect the company's reputation and be ethical (Kim and Lee, 2018).

Business Communications students can critically explore how socially responsible corporations can communicate in a manner that reflects corporate values consistent with corporate social responsibility values (Schmeltz, 2014).

In addition, business students must be aware of acceptable behavior and the dangers that exist related to sexual harassment and sexual violence. The plethora of sexual harassment allegations and court cases have put two worldwide movements and organizations in the spotlight. The *MeToo* movement, an organization founded by Tarana Burke committed to putting a stop to sexual violence. The *Time's Up* movement, an organization founded by Christy Hubegger that is focused on leading changes that result in improved workplace equity through the removal of inequitable power imbalances and safety issues (Langone, 2018). Now more than ever, business communication classes should help students critically explore the impact of their *words and actions* in a business setting and how workplace inequities are perpetuated through formal and informal channels of communications.

Sexual harassment and rape are problems on many college campuses. This year marks the 46th birthday of the landmark civil rights law Title IX passed by the US Congress. Although the law has resulted in women now comprising 56% of America's college students, 48% of all tenure-track positions, and the number of female athletes has increased tenfold, 1/5 of the women on college campuses experience sexual assault and nearly half of the students in middle and high school report sexual harassments (Churches, 2018). In fact, Annie Clark, the co-founder of the nonprofit End Rape on Campus has reported that there are repeat offenders that seek out victims time and again without punishment (Saul, 2017). Some forms of the sexual harassment occur on the internet where rules are not clear and protecting oneself can be tricky. Business communications can provide a platform to discuss the proper use of social media and how to be protected in cyberspace in personal and professional settings.

Economic sustainability includes emergency preparedness and the protection of human safety and quality of living (Collier, 2007). Details of hurricane relief issues in Puerto Rico are noted in Table 1. Besides the hurricane relief debacle in Puerto Rico, several America CEOs recently spoke out against the United States of America's child separations at the Mexican border. Students in business communication can learn lessons in crisis communication and corporate social responsibility (CSR) (Chaudri, 2014).

In addition to computer hacking, sexual harassment and other human rights issues, there is also bullying in workplaces and educational institutions. Workplace bullying can include offensive, intimidating, malicious and/or overbearing supervision, criticism, exclusion, work and threats (Landau 2017). Cyber bullying accounts for significant increase in suicides among young people (Twenge, 2017). Business communication classes can provide an opportunity to discuss safe use of the internet even when seeking employment and how surfing the "dark web" or "friending" can lead to contact with unsavory characters who might appear to be "business professionals."

## Method

The method used in this study is a literature review on current pedagogical practices involving the use of current events in education relevant to preparing millennials for professional business communications. The identification of gaps between the literature and current practice should stimulate development of a new curriculum that infuses these critical elements to provide consistency in messaging regarding the serious nature and responsibility of every student as a global citizen. The goal is to embrace these approaches as a way of thinking about teaching and to advocate for incorporating them wherever appropriate.

**Literature Review**

The attempt here is to provide a review of contemporary issues that can be explored critically in a business communications classroom. Exercises can then be developed to provide opportunities for problem-solving to promote responsible digital citizenship, environmental sustainability and social responsibility within a business communications curriculum.

**About Millennials**

With the growth of millennials in the employee base, millennials are at the forefront of many business conversations (Stewart, 2017):

- With regard to duty, millennials do not believe that productivity should be measured by the number of hours worked, but rather by performance output (PricewaterhouseCoopers, 2013).
- With regard to drive, millennials prefer synergistic decision making team environments; (Stewart, 2017).
- Young adults want to feel personally connected to their employers' goals (Raytheon, 2017).
- Millennials do not conceptually link workplace culture with organizational commitment. Their views of duty, work drive, and rewards differ from other generations recommended differences (Stewart, 2017).

This is relevant because as educators prepare millennials for the workforce, it might be helpful for educators to know how the millennials' performance might be evaluated. Stewart (2017) has suggested that to alleviate frustrations between members of varying generations, training is necessary to develop adequate communication strategies that alleviate misconceptions related to the value of duty, responsibility and work obligations. This should include social responsibilities in business contexts.

**Digital Citizenship**

Digital citizenship is associated with how individuals use their power to process their social decision making (Simsek, 2013). The phrase "cyberspace literacy" is used in information technology to refer to the goal of every cyberspace user behaving and participating in a manner that is independent, cultured and critical (Area, 2012). With the use of online social media platforms such as Facebook, Twitter, Snapchat, Instagram and LinkedIn, there is a need for the development of new cyberspace literacy that embraces Web 2.0. Web 2.0 refers to the second generation of the world wide web which is characterized by more dynamic and interactive collaborative and shared web experiences (Wolcott, 2007). Today's emphasis is on the Internet of Things (IOT) which is defined as the giant network of any device (e.g. cell phones and appliances) connected to the Internet (Morgan, 2014). The key issues in the interaction of new literacies and digital citizenship include professional online participation, the exercise of citizen rights, adequate development of technical know-how, formulation and enforcement of values and norms, proper ways to assess the information overload, and the development of critical attitudes (Simsek, 2013). This is particularly important because the terms of use in the fine print that social media and other digital application providers tout, in order to nudge user behavior, is noncommunicative since most users ignore them.

Although millennials believe that cybersecurity is important, their behavior puts themselves and their employers at risk. The types of actions in which millennials engage put themselves and their employers at risk include: failure to update their applications, failure to use two-step authentications, clinking on links regardless of the certainty of their legitimacy, carelessness in sharing their personal information online, and failure to change passwords for each of their key accounts (Raytheon, 2014, 2016, 2017). Further, 77% connect online without password protection and 42% share their passwords with non-family members.

"Cybersecurity" is rapidly becoming connected with global politics as we witnessed a voter concern in the 2016 US Presidential election. In partnership with the National Cyber Security Alliance, Raytheon conducted a study entitled *Securing Our Future: Closing the Cybersecurity Talent Gap* which included a global survey of millennials by Zogby Analytics. Millennial respondents were from Australia, Germany, Japan, Jordan, Poland, Qatar, Saudi Arabia, the United Arab Emirates, United Kingdom and the United States. The population consisted of 1000 adult respondents ages 18 to 26 in 2014 and nearly 3400 in 2016 and again in 2017 (Raytheon, 2014, 2016, 2017). For US respondents, 53% said that political candidates position on cybersecurity determined whether the millennial respondent would support the candidate. Millennials are aware of ensuing cyberthreats and are making a strong connection between the political atmosphere and cybersecurity issues that face the nation (Raytheon, 2016). It was concluded that:

> "The ongoing effort to raise awareness among millennials about the issue, combined with near-daily news regarding cyberattacks, has both made today's young adults increasingly aware of and interested in cybersecurit*y* jobs. This increase in informed young adults will also likely have effects beyond the obvious jobs to be filled — it very well could sway the nation's future at the voting booths, as most young Americans said cybersecurity issues would likely affect how they vote in elections" (Raytheon, 2016).

Further, with respect to the 2016 US Presidential election, there is a movement toward American leadership having fewer interventions in foreign policy which concerns some foreign policy scholars as America has been looked to for global assistance (Restad, 2017). Millennials blame cyberattacks on their loss of trust in the electoral system. Thus, with the digital ease of information sharing, millennials need to be made aware of the impact of foreign policy on businesses guided by scholars in an evidence based, objective manner. The lack of trust seems to beg a desire for more transparency.

In a virtual ethnography study of Chinese students on the Chinese social network called *Renren*, researchers found that students speak out online to assert their rights and show a sense of wanting to be responsible to wider societal issues. The researchers called this cyber civic participation in cultural citizenship and in lifestyle politics rather than traditional formal political mechanisms such as discussions about political parties, debates, service organizations or elections (Ke, 2014). This is political activism.

Besides educators, it has also been advocated that social media marketing intermediaries such as Facebook, – be held accountable to assist users in developing an understanding of digital citizenship and pay more attention to the dignity and safety of their users. Although, they handle a number of cyber bullying, harassment and other abuse complaints, the decisions they make are vague and indecisive (Citron, 2011). In 2018, Facebook was in the news ; and Facebook's founder Mark Zuckerberg was asked to answer to the United States Congress regarding their user privacy practices and transparency issues (Carlson, 2018; Romm, 2018).

According to Mike Ribble, a forerunner in the area of digital citizenship education, digital citizenship should be embraced by educators as a way of thinking and should be incorporated in any curriculum. He advocates the infusion of the "Nine Elements of Digital Citizenship" in the curriculum as follows:

> 1) Digital Access- these lessons address issues related to access to the internet;
> 2) Digital Commerce – these lessons address the buying and selling of goods online safely;
> 3) Digital Communication – these lessons address sharing information online properly and safely;
> 4) Digital Literacy – these lessons address ongoing education on how to use current digital technologies;
> 5) Digital Etiquette – these lessons address the use of technology by following a respectable code of conduct;
> 6) Digital Law- these lessons address the lawful use of technology content found online;
> 7) Digital Rights and Responsibilities – these lessons address the freedom and responsibility of using the internet;
> 8) Digital Wellness – these lessons address how to achieve a balance that promotes physical and psychological well-being;
> 9) Digital Security – these lessons address strategies to protect your safety online (Ribble, 2011).

This framework serves to assist educators in instilling a "consciousness" among students and sense of right and wrong in order to use digital technology in a socially responsible and acceptable manner. Part of the curriculum can include a survey developed to allow students to self-report by assessing their digital etiquette, security and responsible attitudes and behaviors (Nordin, 2016).

Other digital citizenship assessment scales have been developed to advance this area of scholarship. One includes five (5) characteristics: (1) ethics, (2) fluency, (3) rational, reasonable activities, (4) establishing self-identity, and (5) engagement. This five-factor Digital Citizenship Scale is called the S.A.F.E Model, meaning leading character of Self-identity in digital environment, Activity in online (reasonable activity and social/cultural engagement), Fluency for the digital tools, and Ethics for digital environment. The goal of using this instrument is to encourage students to behave as ethical digital citizens (M. C. Kim, Dongyeon 2018). Another scale is the Digital Citizenship Scale which was validated as a valid measure of digital citizenship (Isman, 2014).

Another research team developed a self-report digital citizenship scale used in a survey of 508 American university students which yielded a five-factor structure of digital citizenship. This included: internet "political activism," technical skills, local/global awareness, critical perspective, and network agency (Choi, 2017). The advancements in this area of scholarship is relevant to educators and to the development of a pedagogical framework because it is important for educators, especially in higher education, to remain current in disciplines that they promote and teach. Educators can also make use of these instruments in studying their students' behavior and in their research.

Lessons in digital citizenship in business communications can also include analyzing the impact of technology in a professional setting; considering the ethical use of employer resources; and acknowledging the personal responsibility and dangers inherent in using the technology. Today's professionals must evaluate the appropriateness of online resources and should be educated on the damage that can impact an employer's reputation by a single digital user.

**Environmental and Social Citizenship**

There are several environmental crises looming in America. One of the most egregious is the Flint, Michigan water crisis. Businesses and private citizens are all adversely impacted. In 2014 and 2015, the residents of Genesee County, MI, endured the third largest recorded disease outbreak in American history. The 87 disease cases coincided with changes in the source and treatment of drinking water in Flint's municipal water system (Zahran, 2018). Lead was found in the water (Dengler, 2017). America is also suffering from joblessness, numerous foreclosed homes, unorganized protests, school shootings, environmental justice issues and what has been called an emotional plague. There is global discourse of the political climate and economy, and a need for debate and dialogue about the cultural and political unconsciousness? (Aronowitz, 2014). Given all this strife in a seemingly irresponsible society, how should millennials be prepared in institutions of higher education to become more socially responsible?

Environmental sustainability can be assisted by empowering students toward a renewed commitment to: 1) perceive dominant ideologies and unmask new ones through systems thinking; 2) challenge power relations; 3) pursue leadership practices for social transformation. "[T]houghtful leaders increasingly recognize that we are not only failing to solve the persistent [environmental] problems we face but are in fact causing them. Today's solutions become tomorrow's problems" (Sterman, 2002). Dominant ideologies hidden behind a cloak of goodwill obscure the reality until it is too late. A perspective that includes system thinking allows individuals to look beyond individual blame to a systems approach that examines the system in which these ideologies are allowed to flourish. Transformational leadership provides a framework embracing power from a socially, environmentally and technically responsible manner.

Since implementing social responsibility programs costs companies, there are understandable budgetary concerns. This is particularly true for environmental sustainability initiatives. For example, in sustainable logistics, there are landed cost, cost to service and cost to cost trade-offs to consider in making such financial investment decisions (Foster, 2016, pp. 316-318). Business communications curriculum can provide special knowledge, skills and abilities to empower millennials to be able to communicate as environmental sustainability stewards and advocates in a diplomatic manner with an understanding and sensitivity toward budgetary concerns. Modern day early career professionals are particularly at risk because they are challenged by their increasing reliance on social media and texting for communications (Hershatter, 2010; Lenhart, 2010) rather than face to face interactions (J.-H. Kim, 2017). The centrality of communication pedagogy and a discipline's content is vital (Morreale, 2017).

**Relevance in the Today's Classroom**
American youth demonstrate a lack of current events knowledge (Knowledge Unlimited, 2015). Academic instructors often fail to relate their syllabi to current events. For example, this is true in the tourism and hospitality discipline (Bernasco, 2017). Higher education faculty are encouraged to integrate current events into their curriculum and augment these events with other scholarly works. For example, it would make sense for a faculty member to not only identify interesting current events, even if difficult to discuss, because most issues will at some point impact student professionally and personally and therefore should be in the repertoire of the faculty's scholarship (Vanderbilt Center for Teaching). Syllabi should not be revised to include units on current events if the teaching strategy is merely to chat with students about news stories without contextualization with relevant scholarly work (Rooks, 2014). In social work and biology, research on the integration of current events into the learning experience have reported positive findings on the impact on student performance (Grise-Owens, 2010; Tinsley, 2016).

Just as transaction lawyers must be well versed in current events since political views and public sentiment affect recently enacted and proposed future legislation that apply to his or her legal transactions, so should business managers (Dean, 2013). Further, modern day business transactions are often global and multicultural. Discussing current events in the classroom is an effective strategy for getting students to begin to deconstruct and understand the complexities of multiculturalism (Deardorff, 2011; Galczynski, 2015; Gordon, 2017). Current events can be selected which have multicultural themes.

**Challenges to Integrating Current Events - Authentic Learning**
Current events can be highly engaging by providing opportunities to examine real business problems requiring problem solving, ethical decision making and critical thinking skills. Current events can be used to achieve constructivism and authenticity (*Authentic Instruction and Online Delivery: Proven principles in higher education*, 2011). Reportedly, the use of current events in economics was first proposed in 1983 (Kelley, 1983; Pomykalski, 2015). With regard to business concepts such as economics, it was discovered that students find it difficult to relate to examples and cases if they are not current, "they have either forgotten about it or they have never heard of it" (Ghosh, 2011). Yet, students are excited and are able to relate better to current events (Ghosh, 2011). In a study of an undergraduate leadership class, students' curriculum included being quizzed on news items in newspapers and on television. Students mostly enjoyed and related to those group discussions (Odom, 2015).

The use of currents events is related to the "real world relevance" component of authentic learning which advocates that students use socialization processes and to learn to use their judgment, patience, ability to synthesize unfamiliar contexts and the flexibility to work across disciplinary and cultural boundaries (Lombardi, 2007a, 2007b; Windham, 2007). A fundamental component of authentic learning experiences is student-centered information inquiry. Current events are examples of content and context of learning readily accepted by students as relevant to his or her needs and deemed by faculty to simulate life beyond the classroom. A specific challenge of library media specialists is to partner with teachers to design learning activities and develop assessments that resemble constructive experiences beyond the school (Callison, 2004; Lamb, 2005).

**Identification of Specific Pedagogical Strategies**

There are several published strategies for integrating current events into the classroom. A model technological pedagogical content knowledge (TPACK) case refers to searching weblogs for current events in world affairs by an instructor and requiring students to keep their own blogs to improve their writing and reflection (Cox, 2009). A law professor reported use of simulations and role-playing exercises in areas of bank failures and system risk based on real controversies involving recent news stories focused on present day economic problems (Dean, 2013). Although designed for a law school course, it is certainly relevant to business law courses. Simulations, role playing, experimentation and the use of current events offer authentic learning experiences and are commonly used across business curriculum including: management, marketing, business law, accounting, and increasingly in business communications (Lombardi, 2007b).

In some instances, students are asked to compare their lessons to real world news stories that they find. The students are asked to defend their choice in selecting the news story and whether the news report is relevant for the class assignment. Oftentimes current events are very relevant from a business communications perspective providing opportunities for critical thinking and problem-solving that can spark thinking about social responsibility, environmental sustainability and digital citizenship. They can be discussed in class orally or structured for multiple assignments that address various aspects of business communication that can be critically examined including: crisis management, CSR management, leadership communication, ethical communication, persuasive messaging, ethical leadership and communication, power and communication, influence and communication, communication and social media/cybersecurity, etc. (Garretson, 2014; Hollander, 2000).

Students can also be instructed to find multiple news articles regarding a current event and explore deeply the differences between the sources and the ways the stories were reported to explore lessons in ethical communication and corporate values (Scheibe, 2009). This would improve the students' media literacy, information processing and offer a way to enhance business communications skills. Students are more engaged when they take ownership of discussion topics and recognize relevant news stories. For example, students can interview individuals about news stories and make *evidence-based arguments* in reports on the news subject (Sage Publishing, 2016). Having the students write in journals about current events and other experiences are also effective in providing feedback on concepts that the student perhaps does not understand (Bahmani, 2016; Stanton, 2017). Journaling helps students find their own voice and *self-identities*.

Much has been written about ways to teach with current events. For example, the New York Times published "50 Ways to Teach with Current Events" (Gonchar, 2014). *Evidenced based reporting* of truth is important since students should not merely rely on their opinions (Gonchar, 2014; Leo, 2014). In an article of 11 active learning business class activities, the author shared the strategy of leaving off the outcome of current events and having students fill in the blank by evaluating and deciding on what the resolutions could be (Livingston, 2015).

Further, Towson University's Student Life office administers a Collegiate Readership Program and advocates 25 ways to use newspapers in the classroom. They conduct New York Times Talks Luncheons form meaningful conversations including students and faculty from all disciplines; and they promote a civil engagement program which makes use of New York Times Teaching Toolkit ("25 Ideas for Using Newspapers in the College Classroom," Towson University). Western Kentucky University uses current events as a means of ethical awareness in its Senior Assessment Finance 499 course. Students are to discuss current events found in the Wall Street Journal weekly in online discussion boards and then submit an ethics-based essay at the end of the semester (Western Kentucky University, 2017).

Reportedly, America's approach to news literacy was led by Dr. Howard Schneider, Dean of the Stony Brook University School of Journalism. He advocated that non-journalism majors be taught the principles and practices of the press; i.e. the information processing skills of a newspaper reporter. Related to current events, since it is difficult to separate high quality, fact-based journalism from everything else, Dr. Schneider advocated that students become news literate on how to identify news and how to critically analyze it (Fleming, 2015). News media literacy is multi-disciplined and includes civics instruction, cultural studies, values education and the impact of media. The core challenges are to get students familiar with recognizing quality sources of news, learn with stories to read and read the news daily (Siena, 2017).

**Social/Political Engagement**

Literacy is the quality and state of being literate. To be literate is to have knowledge and competence. Rhetoric is the art of speaking and writing effectively. This is what business communications is all about. It is important that students be led "through the reflection that makes rhetoric intentional" (Lynch,

2013). When dealing with global media, faculty need to be proactive in equipping students with critical literacy skills (Carneiro, 2008).

 Further with regard to news media literacy, it is advocated that students need to be taught how to make a difference through creativity, conversation and political knowledge (Gauntlett, 2015). They also need to learn how media (news, social) feel and fit emotionally and intellectually (Gauntlett, 2015).  The politically oriented current events may lead students to think critically about policy issues (Yob, 2013). When it comes to contentious political issues which can be fervently debated, it is important to set ground rules in order to facilitate a safe class environment (McDaniel, 2011; Yale Center for Teaching and Learning).

Lastly, unfortunately there is a steady stream of ethical scandals such as those noted in Table 1 (Weybrecht, 2016). Students need time to practice, reflect, formulate their thoughts, and formulate their questions. In preparing millennials, they need time to practice the ethics skills building so that they become ethical business professionals and excellent, socially responsible global citizens (Weybrecht, 2016).

## Findings

The gaps in the literature related to the use of current events in education relevant to preparing millennials for professional business communications include focusing on the students' individual voice and critical exploration of issues within the context of business communication. None of the literature reviewed specific to business education emphasized the importance of preparing millennials to **exercise diplomacy** as they forge their business careers. Further, while some research studies advocated that students find their own news stories, there is little scholarship about the 'then what?'  What should instructors do to pull out of students their ability to exercise their individual voice about ethics, policies, legislation, and business practices. Typically, students merely summarize what the current event news article, blog or social media microblog post states. There needs to be much more research and practice focused on providing outlets for students to: (1) express themselves in online discussions; (2) craft **evidence-based judgments**; and (3) debate on the difficult dialogues related to unethical business practices (Weybrecht, 2016).

The literature review enabled us to identify constructs of importance to include in the proposed pedagogical framework for preparing millennials to be socially responsible citizens. One focus was placed on strategic business management and business policy concepts relevant to social responsibility and digital citizenship in this age of social irresponsibility. These concepts are noted in Table 2. The over-arching emphasis was to search for published research that addressed issues in business communications.

Table 2

Literature Review of Socially Responsible Digital Citizenship - Business Communications Pedagogy

| Literature Review of Social Responsibility Business Communications Pedagogy | |
|---|---|
| **Business management curriculum** | **Social responsibility topics** |
| Marketing | Alternative facts – aka False claims, disinformation |
| | Truth in advertising |
| | Transparency |
| Business Law and Ethics | Corporate social responsibility |
| | Environmental sustainability |
| | Hate speech, harassment, violence, domestic terrorism, racism, sexism, xenophobes, homophobes |
| | *Human rights* including immigration politics |
| | Lying |
| Computer Science and Technology | Digital citizenship |
| | Responsible use of online social networks |
| | Cybersecurity, cyberattacks |
| | Cyberbullying |

While we found some discussion of social responsibility, our sources were greatly limited. Based on our literature review, we find a need for additional socially responsible citizenship education in business communication specifically. We advocate that there is a need for a novel pedagogical framework to assist with the implementation of socially responsible citizenship across the business school curriculum particularly business communications – as well as other business programs. Based on our literature review, we recommend a pedagogical framework for developing business management curricula which has the following ten (10) themes that need to be promoted and implemented by higher education administrative leaders and faculty:

1. **Socially responsive ethics –** digital citizenship and environmental sustainability; social justice awareness using current events in teaching
2. **Self- Identity:** In this age of selfies, faculty should *encourage students to establish* their own authentic self-identities.
3. **Diplomat Business Communication –** encourage *the development of experiential learning activities focused on civic participation for interventionist engagement*
4. **Transparency** - the business schools need to begin at home with discussing digital citizen related decisions made by the university and their college related to cyber bullying, harassment, unethical behavior; and with related current events.
5. **Evidenced –based assignments:** *Evidence-based assignments will thwart any inclination for students to indulge in fake news and alt facts. Require* Oral and Written Communication based on sound researched evidence.
6. **Assert their Rights and Opinions –** in concert with discovering their self-identities, encourage students to learn what their rights and opinions are and exercise their voice

7. **Protect human rights** - identify how they can actively participate in human rights protection locally and globally
8. **Political activism** – encourage students to be politically active cyber civic participants
9. **Cyber secure technical skills** – ensure that all business students understand and apply cyber security safeguards
10. **Critical thinking** – It is important that the students develop critical attitudes, critical perspectives and critical thinking skills

## Conclusion

The world is in a moment of experiencing numerous political, ideological, social, economic, cultural, religious, and other crises. This paper in pedagogy provides educators with guidance in the development of curricula that prepares students as the next generation of leaders responsible in the areas of technology, environmental sustainability and social responsibility as global citizens.

The primary recommendation is that in preparing millennial business students, they should receive adequate ethical business communications training in each of their required courses - - i.e. across the business curriculum. As history unfolds, as a matter of interventionist pedagogy, business college students need to be trained to think critically about what it means and what is required of them in serving as an excellent socially responsible digital citizen. Students need to be steered in a positive direction of becoming more actively engaged in the social issues of their time and business communications courses can assist in this effort.

A novel pedagogical framework is offered herein as a solution to address the key issues in the interaction of new literacies and digital citizenship include professional online participation, the exercise of citizen rights, adequate development of technical know-how, formulation and enforcement of values and norms, proper ways to assess the information overload, and the development of critical thinking throughout business curricula.

The implication of this study in pedagogy is that it will provide educators with much needed guidance in the development of curriculums that address the problem area of social responsibility and better prepare students to perform professionally in their careers as excellent digital citizens and communicate both their environmental sustainability concerns with diplomacy and in a manner that is sensitive to their employers' concerns.

**BARBARA BURGESS-WILKERSON** is Professor of Management and SPD Director at Winthrop University. She received a Ph.D. in Higher Education Administration from the University of Pittsburgh and was an OD Consultant and CEO of BBA Consulting Services for 20 years. Her research interests include: Leadership and Business Communication, Emotional Intelligence and Student Professional Development.

**CLOVIA HAMILTON** has a Ph.D. in Industrial & Systems Engineering with a concentration in Management at the University of Tennessee.  She is experienced and educated in sustainable operations management.  Her research interests include: business law & ethics, technology management, academic entrepreneurship, and university/ federal lab technology transfer operations as novel supply chains.

**CHOLITA POSEY GARRISON** received her Ph.D. in Computer Science from Florida State University, Tallahassee FL. She is a Professor of Computer Science at Winthrop University, Rock Hill SC. She worked for 20 years in software development. Dr. Garrison's research areas include: computer security of the user, promoting Computer Science, Software Engineering and ethical implications of technology.

**KEITH ROBBINS** is Chair of the Department of Management and Marketing and Professor of Management in the CBA at Winthrop University. He has authored numerous articles and cases primarily focused on business turnaround strategies.  Keith received his Ph.D. in Strategic Management from the University of South Carolina.